\documentclass[11pt,epsf,letterpaper]{article}%
\usepackage{setspace}
\usepackage{color}
\usepackage{amsfonts,amssymb,amsmath,amsthm}
\usepackage{verbatim}
\usepackage{graphicx}
\usepackage{amsmath}
\usepackage{amsfonts}
\usepackage{amssymb}
\usepackage{color}%
\providecommand{\U}[1]{\protect\rule{.1in}{.1in}}

\begin{document}

\title{Gribov gap equation at finite temperature}
\author{Fabrizio Canfora$^{1,2}$ Pablo Pais$^{1,2}$ Patricio
Salgado-Rebolledo$^{1,3,4}$\\$^{1}$\textit{Centro de Estudios Cient\'{\i}ficos (CECS), Casilla 1469,
Valdivia, Chile.}\\$^{2}$\textit{Universidad Andr\'{e}s Bello, Av. Rep\'{u}blica 440, Santiago,
Chile.}\\$^{3}$\textit{Departamento de F\'{\i}sica, Universidad de Concepci\'{o}n,
Casilla 160-C, Concepci\'{o}n, Chile.}\\$^{4}$\textit{Physique Th\'{e}orique et Math\'{e}matique, Universit\'{e} Libre
de Bruxelles and}\\\textit{International Solvay Insitutes, Campus Plaine C.P. 231, B-1050
Bruxelles, Belgium.}\\{\small canfora@cecs.cl, pais@cecs.cl, pasalgado@udec.cl}}
\maketitle

\begin{abstract}
In this paper the Gribov gap equation at finite temperature is analyzed. The
solutions of the gap equation (which depend explicitly on the temperature)
determine the structure of the gluon propagator within the semi-classical
Gribov approach. The present analysis is consistent with the standard
confinement scenario for low temperatures, while for high enough temperatures,
deconfinement takes place and a free gluon propagator is obtained. An
intermediate regime in between the confined and free phases can be read off
from the resulting gluon propagator, which appears to be closely related to
partial deconfinement.

\end{abstract}

\newpage

\section{Introduction}

\label{introduction}

\qquad One of the most characteristic features of QCD is asymptotic freedom
\cite{Gross-Wilczek,Politzer}, which allows one to perform the standard
perturbative analysis in the ultraviolet regime. On the other hand, the
infrared regime of the theory is not well understood yet from the analytical
point of view, as the running coupling is large for low energies.\ Indeed,
color confinement is one of the main open problems in theoretical physics.

The standard perturbative approach to avoid overcounting of gauge equivalent
configurations in Yang-Mills (YM) theory is to introduce a gauge fixing
condition in the functional integral (the Landau gauge will be considered in
the following). However, as Gribov pointed out \cite{Gri78}, the Landau gauge
condition does not fix the gauge completely. Shortly after, Singer showed
that, due to the non-trivial nature of the fiber bundle structure of
YM-theory, any true gauge condition presents this obstruction \cite{singer}
(see also \cite{jackiw}). The presence of Gribov copies close to the identity
induces the existence of non-trivial zero modes of the Faddeev-Popov operator,
which make the path integral ill defined. Even when perturbation theory around
vacuum is not affected by Gribov ambiguity when YM-theory is defined over a
flat space-time\footnote{In the curved case, the pattern of appearance of
Gribov copies can be considerably more complicated: see in particular
\cite{P1,P2,P3,P4}. Therefore, only the flat case will be considered.} with
trivial topology \cite{SS05}, Gribov copies have to be taken into account when
considering more general cases \cite{hedgehog} or when non-perturbative
phenomena are studied.

The most effective method to eliminate Gribov copies (proposed by Gribov
himself in \cite{Gri78} and refined in \cite{DZ89,Zwanziger-Ren}) corresponds
to restricting the path integral to the so-called Gribov region, which is the
region in the functional space of gauge potentials over which the
Faddeev-Popov operator is positive definite. In \cite{DZ89} Dell'Antonio and
Zwanziger showed that all the orbits of the theory intersect the Gribov
region, indicating that no physical information is lost when implementing this
restriction. Even though this region still contains copies with non-trivial
winding number \cite{baal}, this restriction has remarkable effects. In fact,
the gluon propagator is suppressed in the infrared and the ghost propagator is
enhanced, which has opened a way to understand color confinement
\cite{SS05,VZ}. A local and renormalizable effective action for YM-theory
whose dynamics is restricted to the Gribov horizon and that yields the same
results for the field propagators was constructed in \cite{Zwanziger-Ren,
local1, local2,local3,local4} by adding extra fields to the action. Later, an
improved action was proposed by considering suitable condensates, which leads
to propagators and glueball masses in agreement with the lattice data
\cite{Sorella:2011,Dudal:2008, sorellaPRL}. With the same action, one can also
solve the old problem of the Casimir energy in the MIT-bag model
\cite{fabrizio}.

Even though it is an experimental fact that quarks and gluons are confined and
color charged states are unobservable as asymptotic states at low
temperatures, it is expected that at high temperatures ($T_{c}\sim150-200$
MeV) they become free \cite{le bellac,yagi}. Such a phase transition from
confinement to quark-gluon plasma (QGP) should be described within the
framework of finite-temperature field theory allowing a better understanding
of natural scenarios as the early universe or compact star physics
\cite{yagi,Glendenning,kapusta2}. The high-temperature sector for the theory
corresponds to the perturbative region, in which gluons are physical states.
In this regime it is necessary to include thermal loop corrections, which may
yield a dynamical thermal mass generation for the gauge fields \cite{yagi}. In
particular, the hard thermal loop approximation cancels infrared divergences
coming from the Matsubara frequencies, allowing one to study plasma
oscillations. On the other hand, lattice QCD allows one to handle
non-perturbative phenomena at finite temperature, such as phase transitions.
The critical temperature for QGP has been subject of several studies as well
as its relation with the energy scale in the context of quenched QCD
\cite{Korthals Altes,gupta}. Moreover, some recent analyses (see, in
particular, \cite{Liao:2007,Pisarski:2009,Hidaka:2008,Kashiwa:2013} and
references therein) strongly support the existence of an intermediate regime
which lies in between the confined phase and the free phase. Within this
regime, some features of the confined phase coexist with the high-temperature
plasma phase. Although it is not clear yet whether this intermediate phase
corresponds to a phase transition or to a cross-over, it can be safely assumed
that such an intermediate regime does appear.

In this paper we will analyze the semi-classical Gribov approach to QCD at
finite temperature\footnote{A remark on the terminology: in the following we
will denote by \textquotedblleft critical temperatures\textquotedblright\ the
temperatures which correspond to changes of the qualitative behavior of the
solutions of the finite-temperature Gribov gap equation. Although the present
analysis by itself is not enough to prove rigorously the appearance of a phase
transition (since suitable order parameters should be identified and
analyzed), we think that this terminology is useful to emphasize the sharp
differences in the behavior of the Gribov gluon propagator as the temperature
changes.} extending the pioneering works \cite{zwanziger, fukushima}. To this
aim, the finite-temperature theory at one loop will be restricted to the
Gribov region and the existence of phase transitions from confinement to gluon
plasma will be analyzed. Since gluon deconfinement is associated to the
presence of dynamical thermal mass \cite{arnold}, the contribution of thermal
loops to the finite-temperature analysis cannot be ignored when implementing
this restriction. The semi-classical Gribov analysis shed considerable light
on the non-perturbative behavior of the gluon propagator. Indeed, at zero
temperature, the existence of a non-trivial solution to the Gribov gap
equation implies that the gluon propagator has imaginary poles and
consequently gluons do not belong to the physical spectrum. Hence, there are
two important requirements that the finite-temperature gap equation must
satisfy. Firstly, the finite-temperature gap equation should have, when the
temperature is low enough, solutions close to the zero-temperature one,
describing confined gluons. Secondly, when the temperature is high enough, the
finite-temperature gap equation should describe propagating gluons. Although
it is not easy to satisfy both conditions \cite{zwanziger,fukushima}, here we
will show that not only they can be satisfied, but also that the
finite-temperature gap equation discloses the presence of a new regime in
between the confined and free regimes, which appears to be closely related to
the intermediate regime mentioned before.

The paper is organized as follows: in Section
\ref{semi-classical-gribov-approach-qcd} the quantization of YM-theory and the
semi-classical Gribov approach at zero temperature are briefly reviewed
following the lines of \cite{SS05}. In Section
\ref{finite-temperature-analysis} the main considerations for the one-loop
finite temperature analysis are exposed, and a thermal gap equation is
derived. Section \ref{phase_transition} is devoted to the numerical study of
phase transition taking into account different possible temperature dependence
for the QCD running coupling. Finally, in Section \ref{discussion}, we discuss
the results and compare them with the known literature. \newline

Note added: recently, Ref. \cite{reinosa} was posted on arXiv, where this
problem is also studied by a different approach.

\section{Semi-classical Gribov approach to QCD}

\label{semi-classical-gribov-approach-qcd}

\qquad In this section we will briefly review the semi-classical procedure to
restrict the path integral formulation of YM-theory to the Gribov region
following the lines of \cite{Gri78,SS05}.

The action functional for $SU(N)$ Euclidean YM-theory is given by%
\begin{equation}
S_{EYM}\left[  A\right]  =\frac{1}{4g_{0}^{2}}\int d^{4}xF_{\mu\nu}^{a}%
F_{a}^{\mu\nu},\label{se}%
\end{equation}
where $g_{0}$ is the coupling constant, $F_{\mu\nu}^{a}=\partial_{\mu}A_{\nu
}^{a}-\partial_{\nu}A_{\mu}^{a}+f^{a}{}_{bc}A_{\mu}^{b}A_{\nu}^{c}$ is the
field strength tensor associated to the four-potential $A_{\mu}=A_{\mu}%
^{a}T_{a}$ and $\left\{  T_{a}\right\}  $ are the anti-hermitian generators of
the $su(N)$ algebra%
\[
\left[  T_{a},T_{b}\right]  =f^{c}{}_{ab}T_{c},
\]
with $f^{a}{}_{bc}$ are the $su(N)$ structure constants. The action (\ref{se})
is invariant under $SU(N)$ gauge transformations%
\[
A_{\mu}\rightarrow A_{\mu}^{\prime}=h^{\dag}\left(  A_{\mu}+\partial_{\mu
}\right)  h\text{ \ },\text{ \ }h\in SU(N).
\]

The quantum theory can be constructed by defining the Feynman path integral.
In order to sum only over inequivalent configurations, a gauge fixing
condition must be implemented via the Faddeev-Popov's trick. In the Landau
gauge $\partial^{\mu}A_{\mu}=0$, the gauge fixed path integral has the
standard form \cite{Peskin}
\begin{equation}
Z=\mathcal{N}\int DADcD\bar{c}\delta\left(  \partial^{\mu}A_{\mu}\right)
\det\left(  \mathcal{M}\right)  \exp\left(  -S_{EYM}\right)  , \label{z1}%
\end{equation}
where $\mathcal{N}$ is a normalization, and $\mathcal{M}$ is the Faddeev-Popov
operator for the Landau gauge condition:%
\begin{equation}
\mathcal{M}^{a}{}_{b}=-\partial^{\mu}\left(  D_{\mu}\right)  ^{a}{}_{b},
\label{fp}%
\end{equation}
with $\left(  D_{\mu}\right)  ^{a}{}_{b}=\delta_{b}^{a}\partial_{\mu}-f^{a}%
{}_{bc}A_{\mu}^{c}$ the covariant derivative in the adjoint representation.

Due to the presence of Gribov copies \cite{Gri78}, however, the expression
(\ref{z1}) is ill defined. To avoid zero modes of the Faddeev-Popov operator
and eliminate copies, Gribov proposed to restrict the path integral to the
so-called Gribov region $C_{0}$, which corresponds to the region in the
functional space of gauge potentials over which the Faddeev-Popov operator is
positive definite,%
\begin{equation}
C_{0}\equiv\left\{  \left.  A_{\mu},\partial^{\mu}A_{\mu}=0\right\vert
\det\mathcal{M}>0\right\}  .\label{C0}%
\end{equation}
The restriction of (\ref{z1}) to the Gribov region can be implemented by
redefining the generating functional as
\begin{equation}
Z_{G}=\mathcal{N}\int DAD\bar{c}Dc\delta\left(  \partial^{\mu}A_{\mu}\right)
\det\left(  \mathcal{M}\right)  \exp\left(  -S_{YM}\right)  \mathcal{V}\left(
C_{0}\right)  ,\label{z2}%
\end{equation}
where the factor $\mathcal{V}\left(  C_{0}\right)  $ ensures that the
integration is performed only over $C_{0}$. In order to characterize
$\mathcal{V}\left(  C_{0}\right)  $, we look at the connected two-point ghost
function generated by (\ref{z1}):
\begin{equation}
\left\langle \bar{c}^{a}\left(  x\right)  c^{b}\left(  y\right)  \right\rangle
=\mathcal{N}\int DA\delta\left(  \partial^{\mu}A_{\mu}\right)  \exp\left(
-S_{YM}\right)  \det\left(  \mathcal{M}\right)  \left(  \mathcal{M}%
^{-1}\left(  x,y\right)  \right)  ^{ab}.\label{gt}%
\end{equation}
Singularities in (\ref{gt}) correspond to zero modes of the Faddeev-Popov
operator, i.e. infinitesimal Gribov copies. In the momentum representation,
singularities different from $k^{2}=0$ imply that $\mathcal{M}\left(
x,y\right)  $ can become negative definite, and therefore it is evaluated
outside the Gribov horizon. The factor $\mathcal{V}\left(  C_{0}\right)  $
must be such that this kind of singularities is not present. This is known as
the \textquotedblleft no-pole condition\textquotedblright.

The standard connected ghost two-point function (\ref{gt}) can be put in the
form%
\begin{equation}
\left\langle \bar{c}^{a}\left(  x\right)  c^{b}\left(  y\right)  \right\rangle
=\mathcal{N}\int DADcD\bar{c}\delta\left(  \partial^{\mu}A_{\mu}\right)
\exp\left(  -S_{YM}\right)  \langle\overline{c}^{a}(x)c^{b}(y)\rangle
_{A},\label{cc1}%
\end{equation}
with $\langle\overline{c}^{a}(x)c^{b}(y)\rangle_{A}$ the connected ghost
two-point function with $A_{\mu}^{a}$ playing the role of an external field.
To second order in perturbation theory this can be written in momentum space
as%
\begin{equation}
\left\langle \bar{c}^{a}c_{a}\right\rangle _{k;A}=\frac{1}{k^{2}}\left(
1+\sigma\left(  k,A\right)  \right)  \approx\frac{1}{k^{2}}\frac{1}%
{(1-\sigma\left(  k,A\right)  )},\label{cc2}%
\end{equation}
where%
\begin{equation}
\sigma(k,A)=\frac{Nk^{\mu}k^{\nu}}{3\left(  N^{2}-1\right)  k^{2}}\frac{1}{V}%
{\displaystyle\sum\limits_{q}}
\frac{A^{a\lambda}(-q)A_{a\lambda}(q)}{\left(  k-q\right)  ^{2}}\left(
\delta_{\mu\nu}-\frac{q_{\mu}q_{\nu}}{q^{2}}\right)  ,\label{sigma}%
\end{equation}
and $V$ stands for the four-dimensional volume of the Euclidean space-time.
Since $A_{\mu}^{a}(-q)A_{a\nu}(q)$ is a decreasing function of $q^{2}$,
$\sigma(k,A)$ decreases as $k^{2}$ increases and the no-pole condition can be
stated as%
\begin{equation}
\sigma(0,A)=\frac{1}{4}\frac{N}{N^{2}-1}\frac{1}{V}%
{\displaystyle\sum\limits_{q}}
\frac{1}{q^{2}}A_{\mu}^{a}(-q)A_{a}^{\mu}(q)<1.\label{no pole}%
\end{equation}
Hence, the factor $\mathcal{V}\left(  C_{0}\right)  $ needed in (\ref{z2}) to
restrict path integrals to the Gribov horizon is given by $\mathcal{V}\left(
C_{0}\right)  =\Theta\left(  1-\sigma(0,A)\right)  $, where $\Theta\left(
x\right)  =\frac{1}{2\pi i}\int_{-i\infty+\varepsilon}^{i\infty+\varepsilon
}d\eta\frac{e^{\eta x}}{\eta}$ is the Heaviside step function. Implementing
this factor in $Z_{G}$, the quadratic part of the path integral in the field
$A_{\mu}$ can be put in the form%
\begin{equation}
Z_{G}^{quad}=\mathcal{N}\int\frac{d\eta}{2\pi i}e^{f\left(  \eta\right)
}\text{\ \ },\text{\ \ }f\left(  \eta\right)  =\eta-\ln\eta-\frac{3}{2}\left(
N^{2}-1\right)  \sum_{q}\ln\left(  q^{2}+\frac{\eta Ng_{0}^{2}}{N^{2}-1}%
\frac{1}{2V}\frac{1}{q^{2}}\right)  .\label{zc}%
\end{equation}
Using the steepest descent (saddle point) method, (\ref{zc}) can be
approximated by $Z_{G}^{quad}\approx e^{f\left(  \eta_{0}\right)  }$, where
$\eta_{0}$ satisfies the minimum condition $f^{\prime}\left(  \eta_{0}\right)
=0$. Defining the Gribov parameter $\gamma^{4}=\frac{\eta_{0}Ng_{0}^{2}}%
{N^{2}-1}\frac{1}{2V}$, the minimum condition leads to the gap equation
\begin{equation}
1-\frac{Ng_{0}^{2}}{\gamma^{4}\left(  N^{2}-1\right)  2V}-\frac{3Ng_{0}^{2}%
}{4V}\sum_{q}\frac{1}{q^{4}+\gamma^{4}}=0.\label{gap_t=0}%
\end{equation}
The solution of this equation in the infinite volume limit $V\rightarrow
\infty$ is given by $\gamma^{2}=\Lambda^{2}e^{-\frac{64\pi^{2}}{3Ng_{0}^{2}}}%
$, where $\Lambda$ is the ultraviolet cutoff, and it leads to a confining
gauge propagator \cite{SS05}%
\begin{equation}
D_{\mu\nu}^{ab}\left(  q\right)  =\delta^{ab}g_{0}^{2}\frac{q^{2}}%
{q^{4}+\gamma^{4}}\left(  \delta_{\mu\nu}-\frac{q_{\mu}q_{\nu}}{q^{2}}\right)
.\label{prop1}%
\end{equation}
For large $q$, (\ref{prop1}) reduces to the standard perturbative result
\cite{Peskin}. In the infrared, however, the gluon propagator is suppressed,
as it displays imaginary poles. In other words, since $D_{\mu\nu}^{ab}\left(
q\right)  $ has a positivity violating K\"{a}ll\'{e}n--Lehmann representation
\cite{Alkofer:2000wg,Peskin}, gluons cannot be considered as part of the
physical spectrum and the propagator (\ref{prop1}) is interpreted as
confining. Replacing (\ref{prop1}) in (\ref{sigma}) leads to the following
behavior for the ghost propagators (\ref{cc2}), in the infrared limit:
\begin{equation}
\left\langle \bar{c}^{a}c_{a}\right\rangle _{q;A}\underset{q\rightarrow
0}{\longrightarrow}\frac{128\pi\gamma^{2}}{3Ng_{0}^{2}}\frac{1}{q^{4}%
},\label{cc3}%
\end{equation}
which means that the ghost propagator is not free-like, but enhanced for
$q\rightarrow0$.

\section{Finite temperature analysis}

\label{finite-temperature-analysis}

\qquad Finite-temperature YM-theory can be studied using the imaginary time
formalism \cite{le bellac,kapusta}, which relates the corresponding quantum
field theory generating functional with a quantum statistical partition
function through a compactification of the temporal coordinate. In this
formalism, the period of the compactified time is associated with the inverse
of the temperature of a thermal bath, and the partition function can be
written as%
\begin{equation}
Z=\int DA\exp\left(  \frac{1}{4g_{0}^{2}}\int_{0}^{\frac{1}{T}}d\tau\int
d^{3}xF_{\mu\nu}^{a}F_{a}^{\mu\nu}\right)  .\label{zt}%
\end{equation}
Since the temporal integration limits $0$ and $T^{-1}$ are identified, when
passing to momentum space, temperature dependent fields are expanded in a
Fourier series over discrete Matsubara frequencies $\omega_{n}$.%
\begin{equation}
\varphi\left(  \tau,\mathbf{x}\right)  =T%
{\displaystyle\sum\limits_{n=-\infty}^{\infty}}
\int\frac{d^{3}q}{\left(  2\pi\right)  ^{3}}e^{-i\left(  \omega_{n}%
\tau+\mathbf{q\cdot x}\right)  }\varphi\left(  \omega_{n},\mathbf{q}\right)
\text{ \ \ },\text{ \ \ }\omega_{n}=2\pi nT.\label{F}%
\end{equation}

\subsection{Dynamical thermal mass}

\qquad When implementing the gauge fixing, the finite-temperature formalism
must be applied to the generating functional (\ref{z1}), where the Euclidean
action has to be written as a local functional for ghost and gauge fields and
perturbation theory can be applied. For gluons, when considering one-loop
corrections, the resumed gauge propagator in the Landau gauge takes the form
\cite{yagi}%
\begin{equation}
D_{\mu\nu}^{ab}\left(  q\right)  =g^{2}\delta^{ab}\left(  \frac{P_{\mu\nu}%
^{T}\left(  q\right)  }{q^{2}+\Pi_{T}\left(  q\right)  }+\frac{P_{\mu\nu}%
^{L}\left(  q\right)  }{q^{2}+\Pi_{L}\left(  q\right)  }\right)  ,\label{prop}%
\end{equation}
where $g$ is the running coupling and%
\begin{align}
P_{\mu\nu}^{T}\left(  q\right)   &  =\delta_{\mu}^{i}\delta_{\nu}^{j}\left(
\delta_{ij}-\frac{q_{i}q_{j}}{\mathbf{q}^{2}}\right)  ,\label{p}\\
P_{\mu\nu}^{L}\left(  q\right)   &  =\delta_{\mu\nu}-\frac{q_{\mu}q_{\nu}%
}{q^{2}}-P_{\mu\nu}^{T}\left(  q\right)  ,\nonumber
\end{align}
are transverse projectors orthogonal to each other, ($P_{\mu\nu}^{T}q^{\nu
}=P_{\mu\nu}^{L}q^{\nu}=0$, $\delta^{\rho\sigma}P_{\mu\rho}^{T}P_{\sigma\nu
}^{L}=0$) and $\Pi_{T}\left(  q\right)  ,$ $\Pi_{L}\left(  q\right)  $ are the
components of the self-energy $\Pi_{\mu\nu}$ along the projectors (\ref{p})%
\begin{equation}
\Pi_{\mu\nu}\left(  q\right)  =P_{\mu\nu}^{T}\left(  q\right)  \Pi_{T}\left(
q\right)  +P_{\mu\nu}^{L}\left(  q\right)  \Pi_{L}\left(  q\right)
.\label{puv}%
\end{equation}
In the plasma region, where $\omega_{n}>>\left\vert \mathbf{q}\right\vert $,
the self-energy components $\Pi_{T}\left(  q\right)  $, $\Pi_{L}\left(
q\right)  $ are given, in the hard thermal loop approximation, by%
\begin{equation}
\Pi_{T}\left(  q\right)  =\Pi_{L}\left(  q\right)  \approx\frac{Ng^{2}T^{2}%
}{9}\label{pp}%
\end{equation}
which means that, in a hot plasma, gauge fields acquire an effective thermal
mass \cite{yagi}%
\begin{equation}
m_{pl}^{2}=\frac{Ng^{2}T^{2}}{9}.\label{m}%
\end{equation}
In this case the gauge propagator \eqref{prop} takes the form%
\begin{equation}
D_{\mu\nu}^{ab}\left(  q\right)  =\frac{g^{2}\delta^{ab}}{q^{2}+m_{pl}^{2}%
}\left(  \delta_{\mu\nu}-\frac{q_{\mu}q_{\nu}}{q^{2}}\right)  .\label{prop0}%
\end{equation}
It is worth noting that ghost fields do not acquire a thermal mass \cite{le
bellac}, which implies that the no-pole condition (\ref{no pole}) has no extra
terms when one-loop corrections are considered. However, the expression for
the gap equation will be modified by the presence of the effective thermal
mass (\ref{m}), as we will see below.

\subsection{Gluon propagator in the presence of dynamical mass}

\label{poles_propagator}

\qquad The effect of a dynamical mass $m$ in the semi-classical Gribov
approach discussed in Section \ref{semi-classical-gribov-approach-qcd} can be
obtained by adding a term of the form $m^{2}A_{\mu}A^{\mu}$ to the quadratic
action in (\ref{zc}). This approach was studied in \cite{sorella dm} and
modifies the gap equation (\ref{gap_t=0}) as%
\begin{equation}
1-\frac{3g^{2}}{\gamma^{4}\left(  N^{2}-1\right)  2V}-\frac{3Ng^{2}}{4V}%
\sum_{q}\frac{1}{q^{4}+m^{2}q^{2}+\gamma^{4}}=0.\label{gap3}%
\end{equation}
The solution of this equation, if it exists, defines a massive (partially)
confining gauge propagator%
\begin{equation}
\bar{D}_{\mu\nu}^{ab}\left(  q\right)  =\delta^{ab}g^{2}\frac{q^{2}}%
{q^{4}+m^{2}q^{2}+\gamma^{4}}\left(  \delta_{\mu\nu}-\frac{q_{\mu}q_{\nu}%
}{q^{2}}\right)  .\label{prop2}%
\end{equation}
The confining character of this propagator relies on the presence of imaginary
poles, which violates positivity of the spectral density function of the
K\"{a}ll\'{e}n--Lehmann representation \cite{Alkofer:2000wg,Peskin},
indicating that it describes non-physical excitations. However, the presence
of a dynamical mass $m$ allows the possibility for the propagator
(\ref{prop2}) to acquire a physical degree of freedom. In fact, the poles of
(\ref{prop2}) are given by%
\begin{equation}
z_{\pm}=\frac{1}{2}\left(  -m^{2}\pm\sqrt{m^{4}-4\gamma^{4}}\right)
\label{poles}%
\end{equation}
Hence, for $m^{2}\geq2\gamma^{2}$ the propagator $\bar{D}_{\mu\nu}^{ab}\left(
q\right)  $ can describe physical particles. Writing (\ref{prop2}) in the
form
\begin{equation}
\bar{D}_{\mu\nu}^{ab}\left(  q\right)  =\delta^{ab}\frac{g^{2}}{\sqrt
{m^{4}-4\gamma^{4}}}\left[  \frac{z_{+}}{\left(  q^{2}-z_{+}\right)  }%
-\frac{z_{-}}{\left(  q^{2}-z_{-}\right)  }\right]  \left(  \delta_{\mu\nu
}-\frac{q_{\mu}q_{\nu}}{q^{2}}\right)  ,\label{dr}%
\end{equation}
we can see that the propagator splits into two terms with opposite residue
sign, indicating that the gluon field $A_{\mu}$ has only one physical degree
of freedom.

In general, if $m$ is a function of some physical parameter, we can
distinguish three scenarios for the behavior of the propagator.

\begin{itemize}
\item For $m^{2}<2\gamma^{2}$ both poles of (\ref{prop2}) are complex,
indicating that there are no propagating gluonic degrees of freedom (confined phase).

\item For $m^{2}\geq2\gamma^{2}$ only one of the two gluonic degrees of
freedom is physical (partially deconfined phase). Hence, if this regime
appears (as will be shown in the following, it does) it shows qualitative
characteristics both of the confined phase and of the deconfined phase.

\item If there is no solution for the gap equation, the only consistent choice
for the Gribov mass parameter is $\gamma=0$, leading to a free gluon
propagator (deconfined phase).
\end{itemize}

In the present case, the effect of the one-loop thermal (\ref{m}) on the
Gribov restriction will be considered by setting $m=m_{pl}\left(  T\right)  $,
and it will be shown that there exist critical temperatures corresponding to
the above three different regimes. It is worth noting that the inclusion of
such a one-loop mass is fundamental in order to obtain these different phases.

\subsection{Thermal gap equation}

\qquad As has been already discussed in the introduction of this manuscript,
two important requirements for the consistency of the analysis are the
following. Firstly, the finite-temperature gap equation should have, when the
temperature is low enough, solutions close to the zero-temperature one,
describing confined gluons. Secondly, when the temperature is high enough, the
gap equation should have no solution, which describes propagating gluons. As
is well known, these conditions are not easy to satisfy
\cite{zwanziger,fukushima}. In the present paper, we will include the one-loop
perturbative corrections both in the running coupling (see Section
\ref{phase_transition}) and in the field propagators (since the crucial role
of the one-loop mass is well known: see \cite{arnold} and references therein).
In order to write down the gap equation for the finite-temperature case, we
apply the prescription (\ref{F}) to (\ref{gap3}) and take the infinite spatial
volume limit
\begin{equation}
\frac{1}{V}\sum_{q}\rightarrow T\sum_{n}\int\frac{d^{3}q}{\left(  2\pi\right)
^{3}}.\label{p2}%
\end{equation}
Finally, replacing the thermal gluon mass (\ref{m}), we obtain the following
thermal gap equation:
\begin{equation}
\frac{3Ng^{2}T}{8\pi^{2}}\sum_{n}\int_{0}^{\Lambda}\frac{r^{2}dr}{\left(
r^{2}+\omega_{n}^{2}\right)  ^{2}+\frac{Ng^{2}T^{2}}{9}\left(  r^{2}%
+\omega_{n}^{2}\right)  +\gamma^{4}}=1,\label{gap4}%
\end{equation}
where we have adopted polar coordinates, integrated over angular variables,
and\ we defined a radial integration limit $\Lambda$, which corresponds to an
ultraviolet cutoff. Let us note that we have neglected the second term of
(\ref{gap3}), as it goes to zero for an infinite spatial volume. Defining the
dimensionless variables
\begin{align}
R &  =\frac{r}{\Lambda}\text{ \ \ },\text{ \ \ }\lambda=\frac{2\pi T}{\Lambda
}\label{var}\\
\theta_{n} &  =\frac{\omega_{n}}{\Lambda}=n\lambda\text{ \ \ },\text{
\ \ }\Gamma=\frac{\gamma}{\Lambda},\nonumber
\end{align}
the thermal gap equation can be rewritten as%
\begin{equation}
\frac{3Ng^{2}\lambda}{16\pi^{3}}\sum_{n}\int_{0}^{1}\frac{R^{2}dR}{\left(
R^{2}+\theta_{n}^{2}\right)  ^{2}+\frac{Ng^{2}\lambda^{2}}{36\pi^{2}}\left(
R^{2}+\theta_{n}^{2}\right)  +\Gamma^{4}}=1.\label{gap5}%
\end{equation}
The sum over all dimensionless Matsubara frequencies $\theta_{n}$ can be
carried out analytically (see Appendix \ref{sum_frequences}), Leading to
\begin{align}
S\left(  R,\lambda,\Gamma\right)   &  =\sum_{n}\frac{1}{\left(  R^{2}%
+\theta_{n}^{2}\right)  ^{2}+\frac{Ng^{2}\lambda^{2}}{36\pi^{2}}\left(
R^{2}+\theta_{n}^{2}\right)  +\Gamma^{4}}\label{s}\\
&  =\frac{\pi}{2\lambda\sqrt{\frac{N^{2}g^{4}\lambda^{4}}{72^{2}\pi^{4}%
}-\Gamma^{4}}}\left(  \tfrac{\coth\left(  \frac{\pi}{\lambda}\sqrt{R^{2}%
+\frac{Ng^{2}\lambda^{2}}{72\pi^{2}}-\sqrt{\frac{N^{2}g^{4}\lambda^{4}}%
{72^{2}\pi^{4}}-\Gamma^{4}}}\right)  }{\sqrt{R^{2}+\frac{Ng^{2}\lambda^{2}%
}{72\pi^{2}}-\sqrt{\frac{N^{2}g^{4}\lambda^{4}}{72^{2}\pi^{4}}-\Gamma^{4}}}%
}-\tfrac{\coth\left(  \frac{\pi}{\lambda}\sqrt{R^{2}+\frac{Ng^{2}\lambda^{2}%
}{72\pi^{2}}+\sqrt{\frac{N^{2}g^{4}\lambda^{4}}{72^{2}\pi^{4}}-\Gamma^{4}}%
}\right)  }{\sqrt{R^{2}+\frac{Ng^{2}\lambda^{2}}{72\pi^{2}}+\sqrt{\frac
{N^{2}g^{4}\lambda^{4}}{72^{2}\pi^{4}}-\Gamma^{4}}}}\right)  ,\nonumber
\end{align}
the gap equation takes the form%
\begin{equation}
\frac{3Ng^{2}\lambda}{16\pi^{3}}\int_{0}^{1}dRR^{2}S\left(  R,\lambda
,\Gamma\right)  =1,\label{gap6}%
\end{equation}
which defines $\gamma$ as a function of $\lambda$%
\begin{equation}
\gamma=\Lambda\Gamma\left(  \lambda\right)  .\label{gamma}%
\end{equation}

\section{The three regimes}

\label{phase_transition}

\qquad As we have shown in Section \ref{finite-temperature-analysis}, the
effective gluon propagator (\ref{prop2}) can lead to three different regimes
for gluons depending on the value of the thermal mass $m_{pl}(T)$, which in
turn depends on the temperature $T$. These three regimes can be associated to
two transition temperatures. In this section we present the numerical analysis
of the gap equation (\ref{gap6}) for QCD ($N=3$) in the high-temperature
regime and subsequently we study a possible infrared continuation.

\subsection{High temperature running coupling}

\begin{figure}[ptb]
\begin{center}
\hspace{-0.2cm} \includegraphics[width=0.7\textwidth,angle=0]{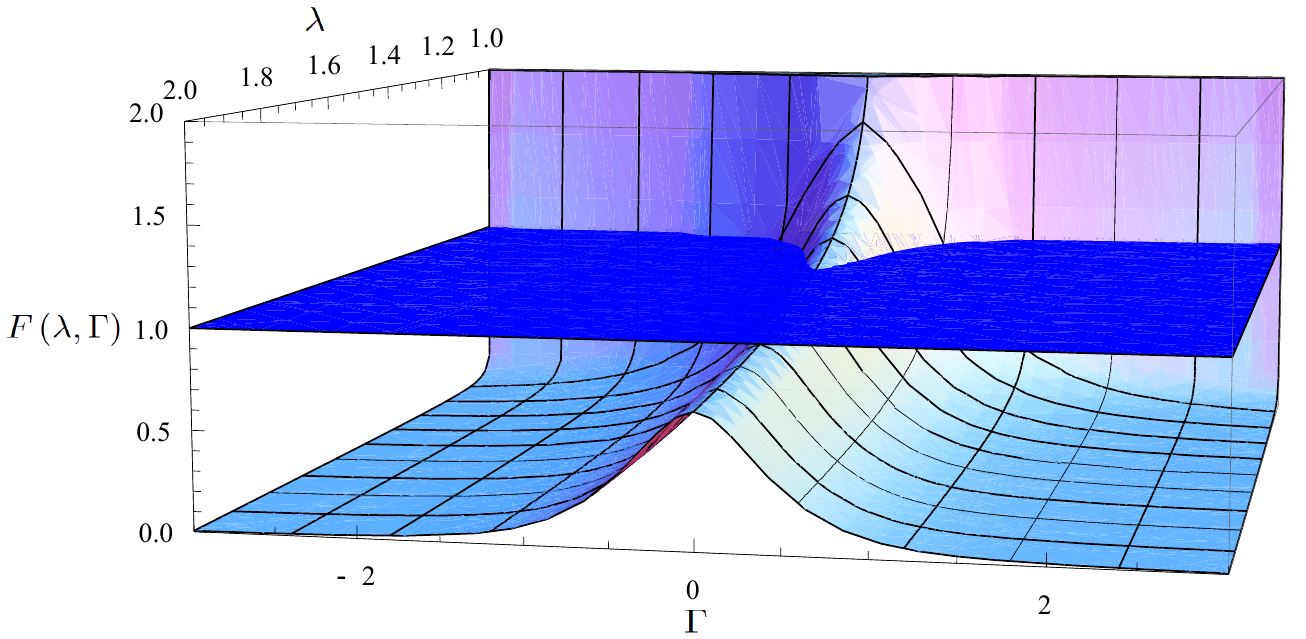}
\end{center}
\caption[3pt]{{\protect\small {Plot of the surface $F$ for different values of
$\lambda$ and $\Gamma$. The intersection with the plane $Y=1$ occurs for
$\lambda$ below the critical value $\lambda_{c}^{\left(  1\right)  }=1.4$.}}}%
\label{caso1a}%
\end{figure}

\begin{figure}[ptb]
\begin{center}
\hspace{0.5cm} \includegraphics[width=0.7\textwidth,angle=0]{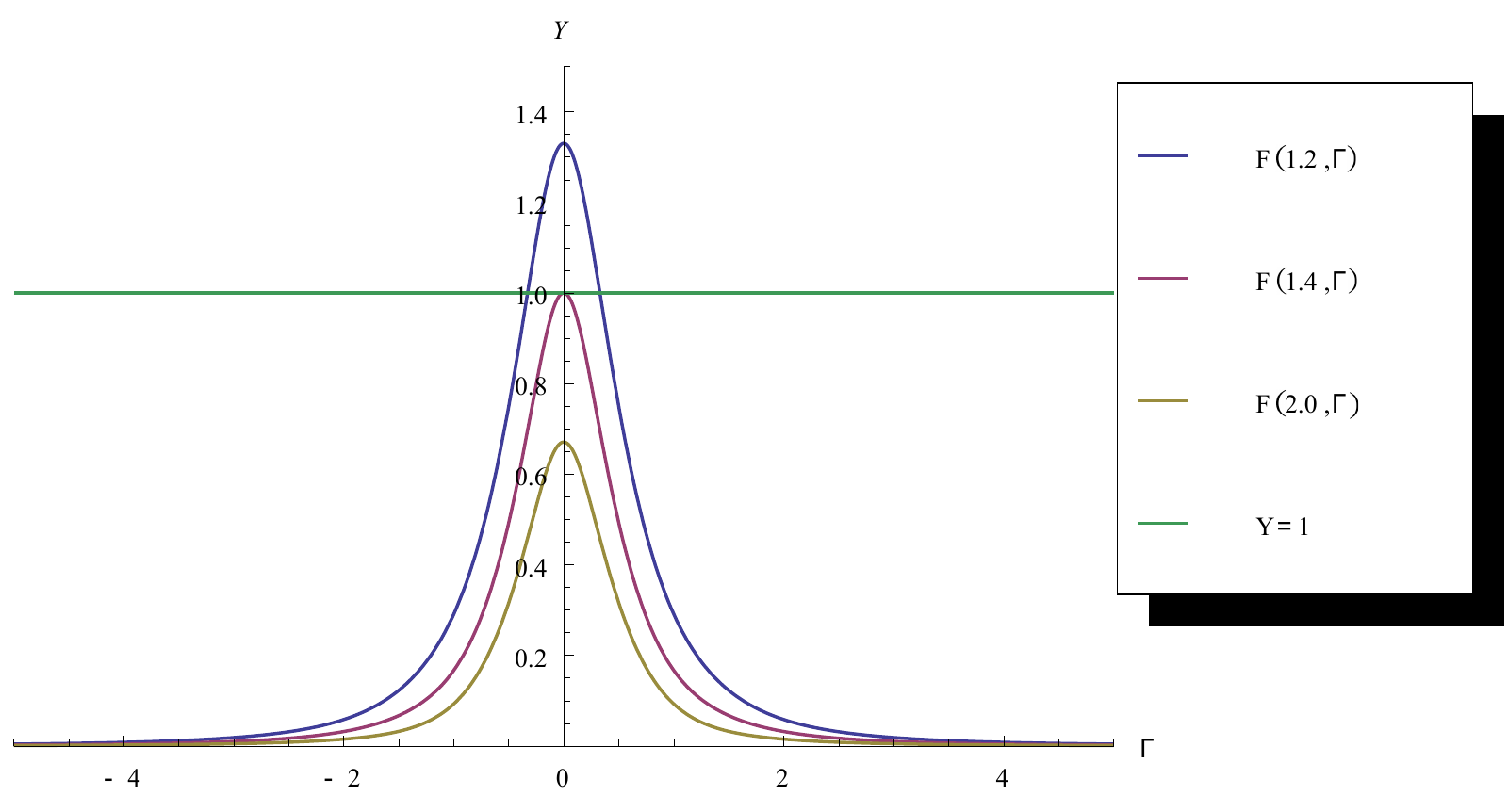}
\end{center}
\caption[3pt]{{\protect\small {Plot of $F$ as a function of $\Gamma$, for
$\lambda=1.2$, $1.4$ and $2.0$. }}}%
\label{caso1b}%
\end{figure}

\begin{figure}[ptb]
\begin{center}
\includegraphics[width=0.55\textwidth,angle=0]{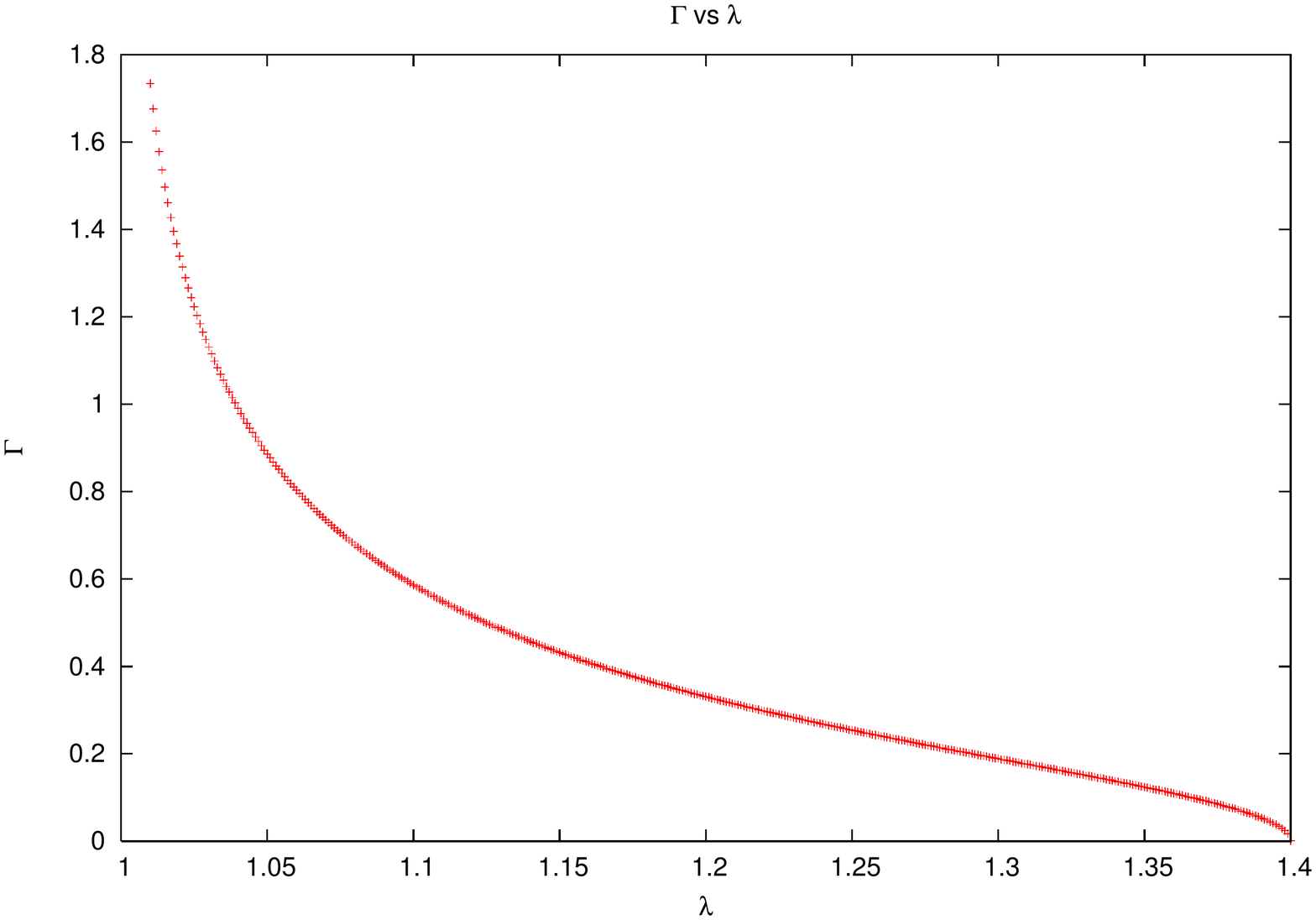}
\end{center}
\caption[3pt]{{\protect\small {Plot $\Gamma$ vs. $\lambda$. At $\lambda
^{\left(  1\right)  }\sim1.4$ there exists a phase transition from a
deconfined phase to a semi-confined one, which corresponds $\frac{T_{c}^{1}%
}{\Lambda}=0.22$.}}}%
\label{gamma_vs_lambda_logaritmica}%
\end{figure}

\qquad Let us consider the thermal gap equation in the limit of high
temperatures $T>>1$. In finite-temperature QCD, the one-loop running coupling
depends on the temperature $T$ (or, in our case, on $\lambda$) as
\cite{zwanziger,huang}
\begin{equation}
g^{2}\left(  \lambda\right)  =\frac{8\pi^{2}}{11\ln\left(  \frac{2\pi
T}{\Lambda_{QCD}}\right)  }=\frac{8\pi^{2}}{11\ln\left(  \alpha\lambda\right)
},\label{g3}%
\end{equation}
where we have defined the ratio between the cutoff $\Lambda$ and the energy
scale $\Lambda_{QCD}$ as
\begin{equation}
\alpha\equiv\frac{\Lambda}{\Lambda_{QCD}}.\label{a}%
\end{equation}
For the left hand side of (\ref{gap6}), we define the function
\begin{equation}
F\left(  \lambda,\Gamma\right)  =\frac{9g^{2}\lambda}{16\pi^{3}}\int_{0}%
^{1}dRR^{2}S\left(  R,\lambda,\Gamma\right)  .\label{f}%
\end{equation}
Then the solution for the gap equation corresponds to the intersection of the
curves $Y=F\left(  \lambda,\Gamma\right)  $ with $Y=1$. In order to obtain the
qualitative behavior for the solutions, we will consider $\alpha=1$ in the
analysis below (as it will be explained later on, the qualitative behavior of
the gluon propagator does not depend on the value of $\alpha$). From Figure
\ref{caso1a}, we see that the existence of solution depends on the
temperature. In fact, the intersection occurs for $\lambda$'s below a critical
value $\lambda_{c}^{\left(  1\right)  }=1.4$, see Figure \ref{caso1b}. This
corresponds to a phase transition at temperature
\begin{equation}
\frac{T_{c}^{\left(  1\right)  }}{\Lambda_{QCD}}=0.22.\label{t1}%
\end{equation}
For $T>T_{c}^{\left(  1\right)  }$ there is no solution for the
gap equation (\ref{gap6}). In this case the only consistent choice for the
Gribov parameter is $\gamma=0$, indicating that this regime represents the
free phase. On the other hand, for $T<T_{c}^{\left(  1\right)  }$, there is a
solution for the gap equation, which define the Gribov parameter $\gamma$.
Therefore, as is shown in Figure \ref{gamma_vs_lambda_logaritmica},
$\Gamma=\gamma/\Lambda$ decreases as $\lambda$ increases and vanishes for
$\lambda^{\left(  1\right)  }=1.4$. \begin{figure}[ptb]
\begin{center}
\includegraphics[width=0.55\textwidth,angle=0]{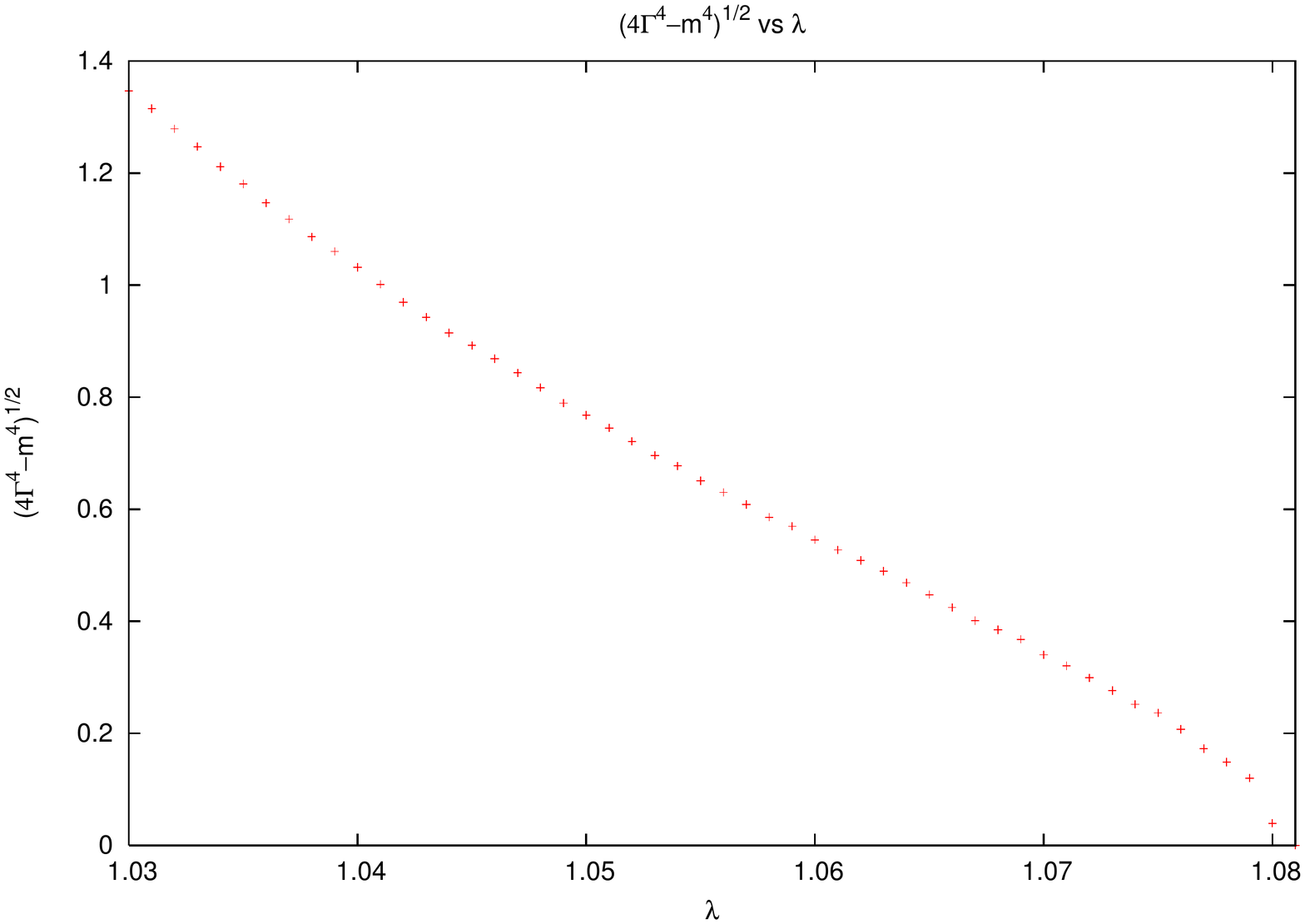}
\end{center}
\caption[3pt]{{\protect\small {Plot $\sqrt{4\Gamma^{4}-m^{4}}$ vs. $\lambda$.
At $\lambda^{(2)}\sim1.08$ there exists a phase transition from a
semi-confined phase to a confined one, which corresponds $\frac{T_{c}^{2}%
}{\Lambda}=0.17$.}}}%
\label{discriminant_vs_lambda_logaritmica}%
\end{figure}Even though for $\lambda<1.4$ there is a solution for the gap
equation, the propagator is still not completely confining. As we saw in
Section \ref{finite-temperature-analysis}, depending on the sign of the
discriminant in (\ref{poles}), a partial or total confinement can take place.
In this case, the change of sign in (\ref{poles}) occurs for $\lambda
_{c}^{\left(  2\right)  }=1.08$ (see Figure
\ref{discriminant_vs_lambda_logaritmica}), which corresponds to
\begin{equation}
\frac{T_{c}^{\left(  2\right)  }}{\Lambda_{QCD}}=0.17.\label{t2}%
\end{equation}
Hence, two phase transitions are found as the temperature decreases: a
deconfined/partially deconfined phase transition at $T_{c}^{\left(  1\right)
}$ and a partially deconfined/confined phase transition at $T_{c}^{\left(
2\right)  }$. In the intermediate phase, only one degree of freedom of the
gluon field is physical, as discussed in Section \ref{poles_propagator}.

\subsection{Infrared continuation}

\begin{figure}[ptb]
\begin{center}
\includegraphics[width=0.7\textwidth,angle=0]{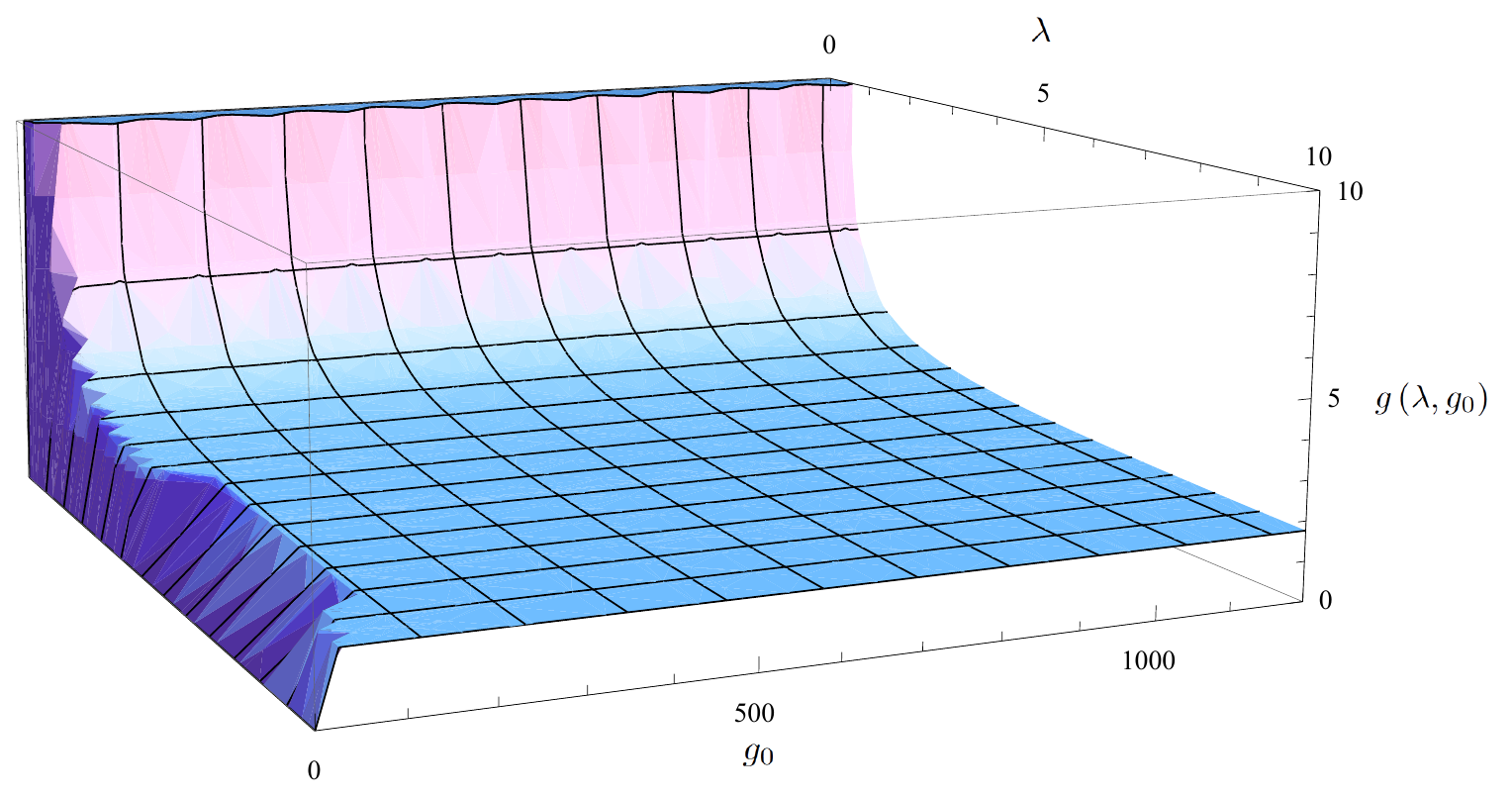}
\end{center}
\caption[3pt]{{\protect\small {Plot of the running coupling $g$ as a function
of $g_{0}$ and $\lambda$. For $g_{0}$ large, $g$ becomes almost insensible to
small variations of $g_{0}$.}}}%
\label{caso2_g}%
\end{figure}

\begin{figure}[ptb]
\begin{center}
\hspace{-0.2cm} \includegraphics[width=0.7\textwidth,angle=0]{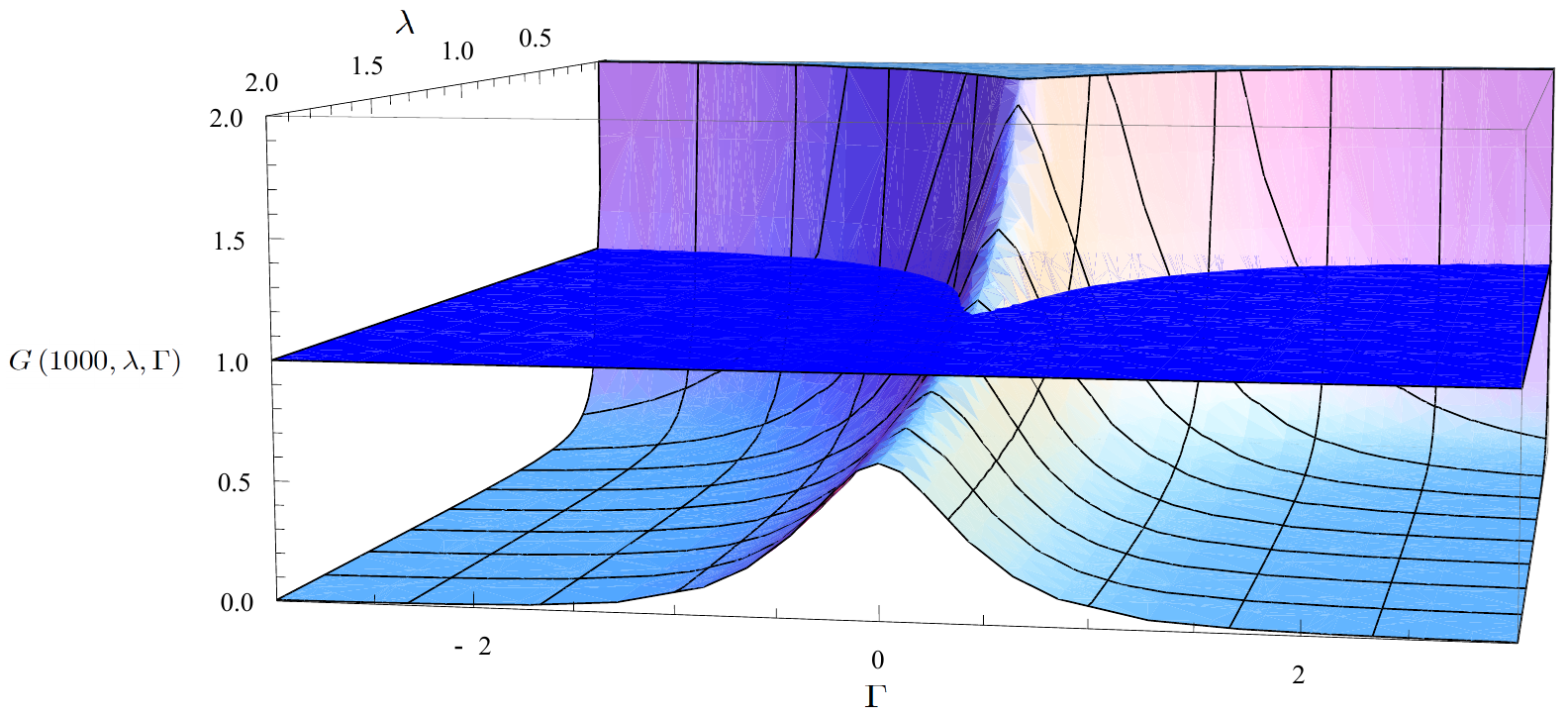}
\end{center}
\caption[3pt]{{\protect\small {Plot of the surface $F$ for different values of
$\lambda$ and $\Gamma$. The intersection with the plane $Y=1$ occurs for
$\lambda$ below a critical value $\lambda_{c}^{\left(  1\right)  }=1.17$.}}}%
\label{caso2a}%
\end{figure}

\begin{figure}[ptb]
\begin{center}
\includegraphics[width=0.45\textwidth,angle=0]{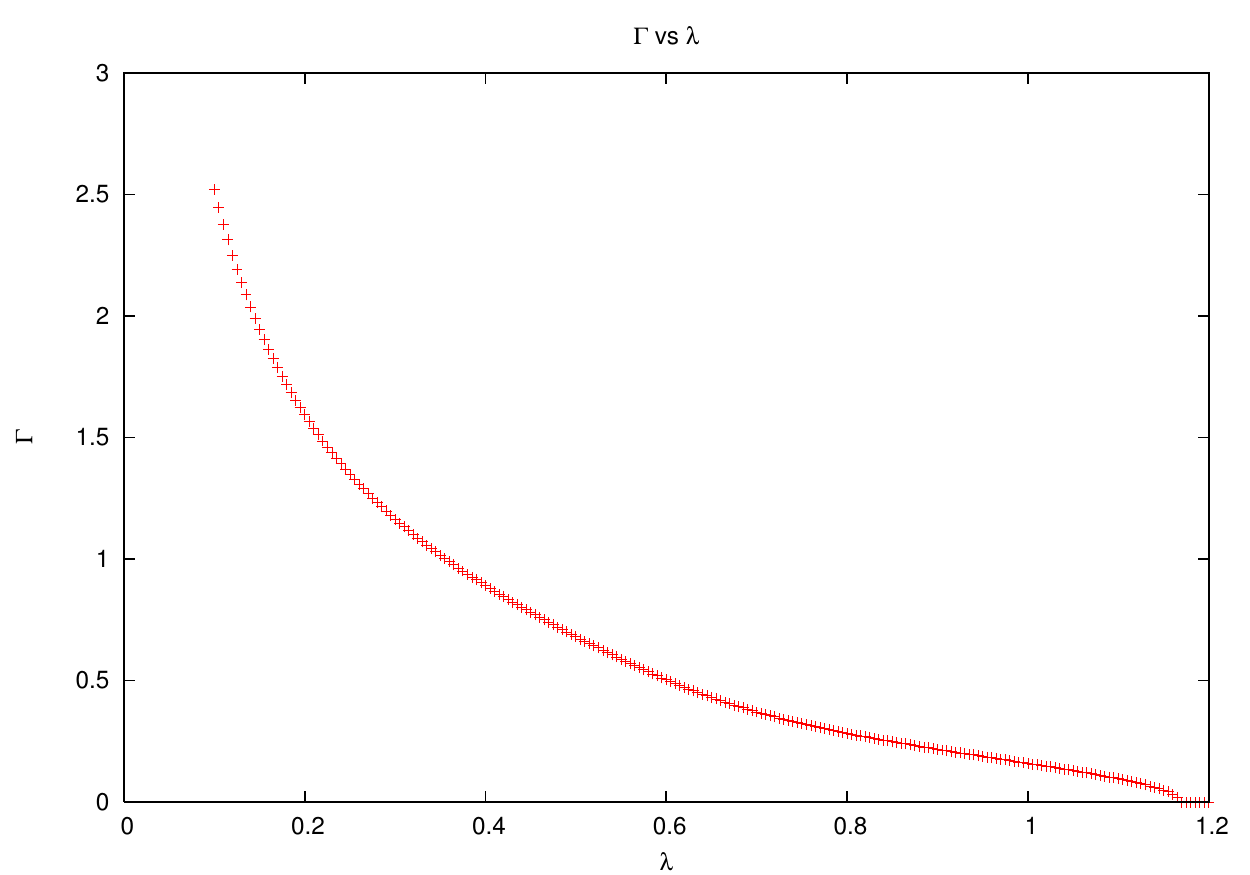}
\end{center}
\caption[3pt]{{\protect\small {Plot $\Gamma$ vs. $\lambda$. At $\lambda
^{\left(  1\right)  }\sim1.17$ there exists a phase transition from a
deconfined phase to a semi-confined one, which corresponds to $\frac{T_{c}^{(1)}%
}{\Lambda}=0.19$.}}}%
\label{gamma_vs_lambda}%
\end{figure}

\begin{figure}[ptb]
\begin{center}
\includegraphics[width=0.45\textwidth,angle=0]{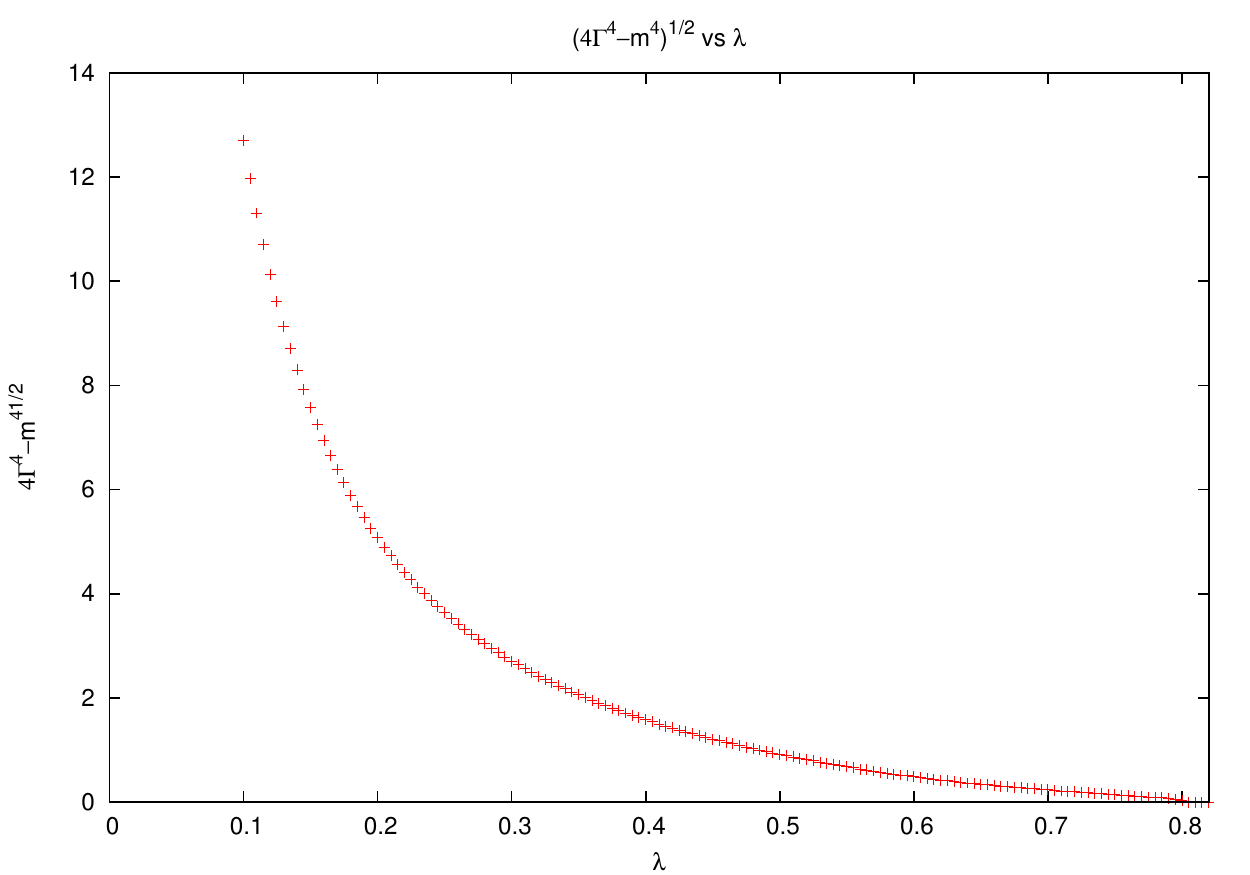}
\end{center}
\caption[3pt]{{\protect\small {Plot $\sqrt{4\Gamma^{4}-m^{4}}$ vs. $\lambda$.
At $\lambda^{(2)}\sim0.81$ there exists a phase transition from a
semi-confined phase to a confined one, which corresponds to $\frac{T_{c}^{(2)}%
}{\Lambda}=0.13$.}}}%
\label{discriminant_vs_lambda}%
\end{figure}

\qquad In order to extend the analysis of the previous subsection to the
low-temperature regime, we need a prescription to extend the definition
(\ref{g3}) for $\lambda<1$. A way to extend the running coupling to the
infrared regime in zero-temperature QCD has been developed in \cite{prosperi}
in the framework of quark-antiquark potentials by adding a non-perturbative
contribution to the Wilson loop. In the finite-temperature case, the analog
extension reads%
\begin{equation}
g^{2}\left(  g_{0},\lambda\right)  =\frac{g_{0}^{2}}{1+\frac{11}{16\pi^{2}%
}g_{0}^{2}\ln\left(  1+\alpha^{2}\lambda^{2}\right)  }.\label{coupling2}%
\end{equation}
This expression reduces to (\ref{g3}) for large $\lambda$ but, in the limit
$\lambda\rightarrow0$ the running coupling reduces to the bare coupling
constant $g_{0}$%
\[
g^{2}\underset{\lambda\rightarrow0}{\longrightarrow}g_{0}^{2}.
\]
This choice is also consistent with the fact that the thermal gluon mass
(\ref{m}) must vanish as $T$ goes to zero%
\[
m_{pl}^{2}\underset{T\rightarrow0}{\longrightarrow}0,
\]
which is a necessary requirement to reduce (\ref{gap3}) to (\ref{gap_t=0}) in
this limit and to connect consistently with the standard $T=0$ results
\cite{SS05}. Let us note that for large $g_{0}$ the behavior of $g\left(
g_{0},\lambda\right)  $ becomes insensible to small variations of $g_{0}$
itself; see Figure \ref{caso2_g}. This is also consistent with the fact that
in quantum field theory bare quantities are infinite but unobservable and they
need to be renormalized. Replacing the expression (\ref{coupling2}) (with $\alpha=1$) in
the gap equation (\ref{gap6}), the left hand side takes the form
\begin{equation}
G\left(  g_{0},\lambda,\Gamma\right)  =\frac{9g^{2}\lambda}{16\pi^{3}}\int
_{0}^{1}dRR^{2}S\left(  R,g_{0,}\lambda,\Gamma\right)  .\label{G}%
\end{equation}
where $S\left(  R,g_{0,}\lambda,\Gamma\right)  $ is obtained replacing
(\ref{coupling2}) in (\ref{s}). Then the solution for the gap equation again
corresponds to the intersection of the curves $Y=G\left(  g_{0},\lambda
,\Gamma\right)  $ and $Y=1$, whose existence depends on $\lambda$ (see Figure
\ref{caso2a}). \begin{figure}[ptb]
\begin{center}
\hspace{0.5cm} \includegraphics[width=0.7\textwidth,angle=0]{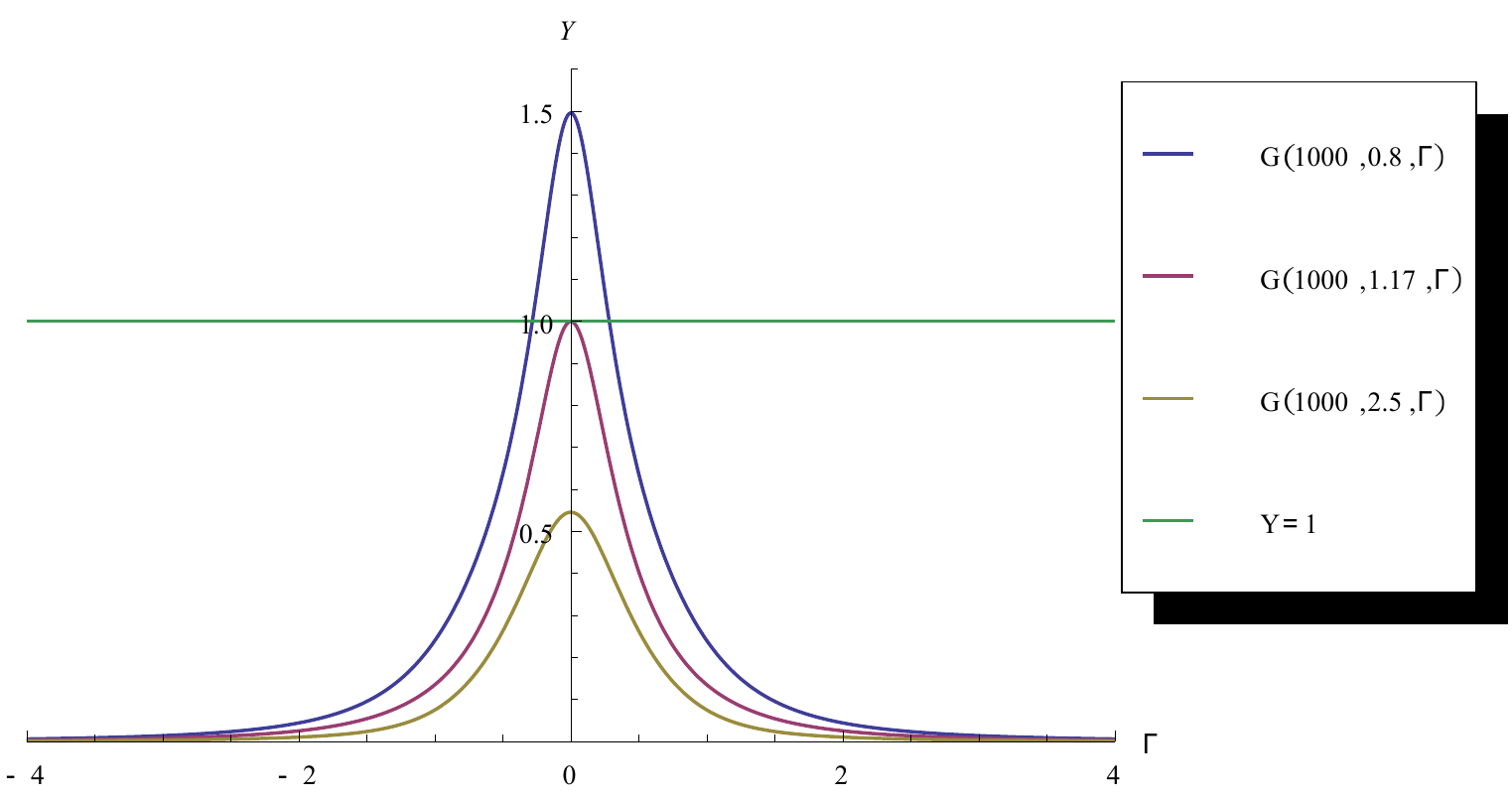}
\end{center}
\caption[3pt]{{\protect\small {Plot of $F$ as a function of $\Gamma$, for
$\lambda=0.8$, $1.17$ and $2.5$.}}}%
\label{caso2b}%
\end{figure}Similarly to the previous subsection, we find two phase
transitions. Choosing $g_{0}=1000$, the deconfined/partially deconfined phase
transition occurs for the critical value $\lambda_{c}^{\left(  1\right)
}=1.17$ (see Figures \ref{caso2b} and \ref{gamma_vs_lambda}), which
corresponds to
\begin{equation}
\frac{T_{c}^{\left(  1\right)  }}{\Lambda_{QCD}}=0.19,\label{t1_ir}%
\end{equation}
while the partially deconfined/confined phase transition now occurs for
$\lambda_{c}^{\left(  2\right)  }=0.81$ (see Figure
\ref{discriminant_vs_lambda}), i.e.,
\begin{equation}
\frac{T_{c}^{\left(  2\right)  }}{\Lambda_{QCD}}=0.13.\label{t2_ir}%
\end{equation}


The results obtained with the prescription (\ref{coupling2}) are very similar
to the ones obtained in the previous subsection. It is important to note that
the qualitative behavior of the solution of the gap equation and the gluon
propagator does not depend on the value of $\alpha$ in the definition
(\ref{a}). As we can see in Table \ref{table_alpha}, the greater the value of
$\alpha$ that we consider in the analysis (i.e. the greater the cutoff
$\Lambda$ compared with QCD scale $\Lambda_{QCD}$), the greater will be the
numerical values for the critical temperatures for the phase transitions.
Hence, the fact that the integration cutoff $\Lambda$ is much higher than the
QCD scale $\Lambda_{QCD}$ implies that the critical temperatures obtained with
this method, when considering a more realistic ratio between this quantities,
will be greater than the values obtained in this section.

\begin{table}[pth]
\begin{center}%
\begin{tabular}
[c]{|l|l|l|l|l|}\hline
\text{ } & \multicolumn{2}{|c|}{$g_{HT}$} & \multicolumn{2}{|c|}{$g_{IC}$%
}\\\hline\hline
$\alpha$ & $\frac{T_{c}^{(1)}}{\Lambda_{QCD}}$ & $\frac{T_{c}^{(2)}}%
{\Lambda_{QCD}}$ & $\frac{T_{c}^{(1)}}{\Lambda_{QCD}}$ & $\frac{T_{c}^{(2)}%
}{\Lambda_{QCD}}$\\\hline
1 & 0.223 & 0.172 & 0.186 & 0.128\\
10 & 0.437 & 0.331 & 0.414 & 0.301\\
100 & 0.758 & 0.558 & 0.752 & 0.540\\\hline
\end{tabular}
\end{center}
\caption{Critical temperatures $\frac{T_{c}^{(1)}}{\Lambda_{QCD}}$ and
$\frac{T_{c}^{(2)}}{\Lambda_{QCD}}$ for different values of $\alpha$. Here,
$g_{HT}$ and $g_{IC}$ correspond to the running coupling at high temperature
\eqref{g3} and its infrared continuation \eqref{coupling2}, respectively.}%
\label{table_alpha}%
\end{table}

On the other hand, in our analysis we have considered only gluon dynamics
(without quarks). In \cite{Korthals Altes,gupta} it has been found that the
value for the energy scale $\Lambda_{QCD}$ that must be considered depends on
the numbers of flavors that are included in the analysis and there have been
found different values for $T_{c}/\Lambda_{QCD}$ depending on these considerations.

\section{Discussion and future developments}

\label{discussion}

\qquad In this paper it has been shown that the semi-classical Gribov approach
applied to finite-temperature YM-theory is consistent with the presence of a
confined/deconfined phase transition. This is reflected in the fact that the
existence of solutions of the Gribov the gap equation depends on the temperature.

A key ingredient for the consistent description of these different regimes is
the inclusion of a mass term in the gluon propagator, which comes from the
one-loop corrections to the theory. Indeed, if the mass term is not taken into
account, there are no critical temperatures at all and one would be left with
confined gluons at all the temperatures. Furthermore, to include one-loop
corrections is consistent with the fact that the thermal mass (\ref{m}) causes
gluon deconfinement \cite{arnold}.

In order to be able to study the low-temperature limit, we have introduced a
modified running coupling $g$, which interpolates between the standard
perturbative result in the ultraviolet regime and a constant (in principle
infinite but unobservable) for the infrared regime. It is worth to note that
this modification has been considered only for consistency, as it allows the
gluon thermal mass to go to zero for low temperature, but the presence of
these phase transitions does not depend on this fact. Indeed, the same
qualitative behavior for the gluon propagator was obtained when considering
the standard one-loop running coupling (\ref{g3}) and, furthermore, it can be
shown that phase transitions are also present if only a constant coupling is
considered in the whole analysis. We stress that when the Gribov
semi-classical method is implemented at zero temperature but with a
non-trivial Higgs field (see \cite{higgs1,higgs2}), the phase diagram turns
out to be very close to the one obtained in the present paper in agreement
with the Fradkin-Shenker theorem \cite{frashe}.

In this paper we have considered the scaling solution, in which the gluon
propagator \eqref{prop1} vanishes and the ghost propagator \eqref{cc3} blows
up as $1/q^{4}$ in the infrared limit $q\rightarrow0$. On the other hand, it
is clear by now that the decoupling solution (where the gluon propagator goes
to a constant in the infrared limit while the ghost propagator has a free-like
behavior) is the relevant one\footnote{In Ref. \cite{kondo} the effect of
Gribov horizon in the Schwinger-Dyson equations has been studied obtaining
both the scaling and the decoupling solution.} \cite{boucaud1,boucaud2}. The
decoupling solution has a strong lattice support
\cite{cucchieri1,bogolusky,cucchieri2,cucchieri3} and can be obtained
analytically within the refined Gribov-Zwanziger theory by including some
condensates \cite{Sorella:2011,Dudal:2008,sorellaPRL}. It would certainly be
of interest to study the refined Gribov-Zwanziger approach at finite
temperature. However, as this theory includes extra ghost fields necessary to
express the action in a local form, the main technical problem when passing to
the finite-temperature formalism is to determine the boundary conditions that
these extra fields must satisfy. This issue is under investigation and we hope
to come back to this point in the future.

An interesting result of this paper is the appearance of an intermediate
regime in between the confined and free regimes, in which only one of the two
gluonic degrees of freedom is physical, while the other one does not belong to
the physical spectrum. In this sense, this new regime captures traces of both
confined and deconfined regimes. Hence, this scenario could be interpreted as
a partial deconfinement or a semi-QGP phase, which has been studied in
\cite{Pisarski:2009,Hidaka:2008,Kashiwa:2013}. Regimes of this kind can appear
when studying QGP by different methods. In fact, in a very interesting paper
\cite{Liao:2007} the phase transition in hot QCD is analyzed in the context of
electrically and magnetically charged quasi-particles, where the confined
regime corresponds to a magnetically dominated and electrically confined
region while the free regime is described by a magnetically strongly
correlated and electrically dominated region. In between these regimes, a
"postconfined" region is found, where electrically charged excitations are
strongly correlated, which can also be interpreted as a partial deconfinement.
It is reassuring that, even though this method is quite different from our
approach, the qualitative results are in agreement with ours, as far as the
presence of an intermediate regime is concerned.

Despite the fact that pure Yang-Mills theory is interesting in itself, the
inclusion of quarks is important in order to obtain a more realistic model.
This point requires a careful analysis since, as has been shown in
\cite{furui,parappilly,burgio,rojas} (see also \cite{sorella_quarks}), the
quark propagator develops complex poles at the non-perturbative level in the
same way as the gluon propagator does after implementing the Gribov
restriction. According to the analysis for a propagator with complex poles
given in Section \ref{poles_propagator}, the fact that the quark and gluon
propagators share this feature in the infrared strongly suggests that there
could exist an intermediate quark regime as well. On the other hand, the
thermodynamics of quark models with complex mass poles have been studied in
\cite{benic} and it would be interesting to follow its lines when adding
quarks to the Gribov-Zwanziger theory. The study of the equation of state for
gluons, both in the semi-classical Gribov approach and the Gribov-Zwanziger
theory, as well as the inclusion of quarks, presents several technical
difficulties at the analytical and numerical level, and they are currently
under investigation.

Another important subject in the understanding of deconfinement, which is also
the aim of a future work, is the order of the phase transition. In order to
formally associate each regime of Section \ref{phase_transition} with a phase
of a gluon plasma and prove rigorously that the critical temperatures that we
have found determine a phase transition, an order parameter must be
introduced. The natural choice for the order parameter in finite-temperature
formalism is the Polyakov loop, and its computation for the intermediate phase
found in Section \ref{poles_propagator} would shed light on its relation with
partial deconfinement and semi-QGP. On the other hand, as has been explained
in \cite{benic}, the Polyakov loop is very useful when dealing with
non-trivial thermodynamics arising in the presence of complex mass poles.

\section{Acknowledgments}

\qquad We wish to thank David Dudal and Silvio Sorella for very useful
comments. This work has been funded by the Fondecyt Grants No. 1120352. The
Centro de Estudios Cient\'{\i}ficos (CECs) is funded by the Chilean Government
through the Centers of Excellence Base Financing Program of CONICYT. P.P
thanks the Physique Th\'{e}orique et Math\'{e}matique group of the
Universit\'{e} Libre de Bruxelles, the International Solvay Institutes and the
Faculty of Mathematics and Physics of Charles University in Prague for the
kind hospitality during the development of this work. P.P. is supported by
grants from Universidad Andr\'{e}s Bello. P. S-R. is supported by grants from
BECAS CHILE, Comisi\'{o}n Nacional de Investigaci\'{o}n Cient\'{\i}fica y
Tecnol\'{o}gica CONICYT and from Universidad de Concepci\'{o}n, Chile.

\appendix

\section{Sum Over Matsubara Frequencies}

\label{sum_frequences}

\setcounter{equation}{0} \numberwithin{equation}{section}

\qquad Let's consider the gap equation (\ref{gap5})%
\[
\frac{3Ng^{2}\lambda}{16\pi^{3}}\sum_{n}\int_{0}^{1}\frac{R^{2}dR}{\left(
R^{2}+\theta_{n}^{2}\right)  ^{2}+\frac{Ng^{2}\lambda^{2}}{36\pi^{2}}\left(
R^{2}+\theta_{n}^{2}\right)  +\Gamma^{4}}=1,
\]
and let's compute the following sum over the dimensionless Matsubara
frequencies $\theta_{n}$%
\begin{equation}
\sum_{n}\frac{1}{\left(  R^{2}+\theta_{n}^{2}\right)  ^{2}+\frac{Ng^{2}%
\lambda^{2}}{36\pi^{2}}\left(  R^{2}+\theta_{n}^{2}\right)  +\Gamma^{4}}%
=\sum_{n}\frac{1}{P\left(  n^{2}\right)  }, \label{a1}%
\end{equation}
where%
\begin{gather}
P\left(  x\right)  =\lambda^{4}\left(  x+a_{-}\right)  \left(  x+a_{+}\right)
,\label{aP}\\
a_{\pm}=\frac{R^{2}}{\lambda^{2}}+\frac{Ng^{2}}{72\pi^{2}}\pm\sqrt{\frac
{N^{2}g^{4}}{72^{2}\pi^{4}}-\frac{\Gamma^{4}}{\lambda^{4}}}. \label{a3}%
\end{gather}
Using algebraic manipulations, we can write (\ref{a1}) as
\begin{equation}
\sum_{n}\frac{1}{P\left(  n^{2}\right)  }=\frac{1}{\lambda^{4}}\frac{1}%
{a_{+}-a_{-}}\sum_{n}\left(  \frac{1}{n^{2}+a_{-}}-\frac{1}{n^{2}+a_{+}%
}\right)  . \label{a4}%
\end{equation}
Then, using the residue theorem applied sum series%
\[
\sum_{n=-\infty}^{\infty}f\left(  z\right)  =-\sum res\left[  \pi\cot\left(
\pi z\right)  f\left(  z\right)  \right]  ,
\]
we obtain for (\ref{a4})
\begin{equation}
\sum_{n}\frac{1}{P\left(  n^{2}\right)  }=\frac{1}{\lambda^{4}}\frac{1}%
{a_{+}-a_{-}}\left(  \frac{\pi\coth\left(  \pi\sqrt{a_{-}}\right)  }%
{\sqrt{a_{-}}}-\frac{\pi\coth\left(  \pi\sqrt{a_{+}}\right)  }{\sqrt{a_{+}}%
}\right)  . \label{a5}%
\end{equation}
Defining $S\left(  R,\lambda,\Gamma\right)  =\sum_{n}\frac{1}{P\left(
n^{2}\right)  }$ and using (\ref{a3}) we obtain (\ref{s})%
\begin{equation}
S\left(  R,\lambda,\Gamma\right)  =\frac{\pi}{2\lambda\sqrt{\frac{N^{2}%
g^{4}\lambda^{4}}{72^{2}\pi^{4}}-\Gamma^{4}}}\left(  \tfrac{\coth\left(
\frac{\pi}{\lambda}\sqrt{R^{2}+\frac{Ng^{2}\lambda^{2}}{72\pi^{2}}-\sqrt
{\frac{N^{2}g^{4}\lambda^{4}}{72^{2}\pi^{4}}-\Gamma^{4}}}\right)  }%
{\sqrt{R^{2}+\frac{Ng^{2}\lambda^{2}}{72\pi^{2}}-\sqrt{\frac{N^{2}g^{4}%
\lambda^{4}}{72^{2}\pi^{4}}-\Gamma^{4}}}}-\tfrac{\coth\left(  \frac{\pi
}{\lambda}\sqrt{R^{2}+\frac{Ng^{2}\lambda^{2}}{72\pi^{2}}+\sqrt{\frac
{N^{2}g^{4}\lambda^{4}}{72^{2}\pi^{4}}-\Gamma^{4}}}\right)  }{\sqrt
{R^{2}+\frac{Ng^{2}\lambda^{2}}{72\pi^{2}}+\sqrt{\frac{N^{2}g^{4}\lambda^{4}%
}{72^{2}\pi^{4}}-\Gamma^{4}}}}\right)  \label{a6}%
\end{equation}
and the thermal gap equation (\ref{gap5}) takes the form%
\[
\frac{3Ng^{2}\lambda}{16\pi^{3}}\int_{0}^{1}dRR^{2}S\left(  R,\lambda
,\Gamma\right)  =1.
\]


\bigskip

\begin{thebibliography}{99}                                                                                               %


\bibitem {Gross-Wilczek}D. J. Gross and F. Wilczek, \textit{Phys. Rev. Lett.}
\textbf{30}, 1343 (1973);

\bibitem {Politzer}H. D. Politzer, \textit{Phys. Rev. Lett.} \textit{30}, 1346 (1973).

\bibitem {Gri78}V.N. Gribov, \textit{Nucl. Phys}. \textbf{B 139} (1978) 1.

\bibitem {singer}I. M. Singer, \textit{Comm. Math. Phys}. \textbf{60} (1978), 7.

\bibitem {jackiw}R. Jackiw, I. Muzinich, C. Rebbi, \textit{Phys.Rev}.
\textbf{D 17} (1978) 1576.

\bibitem {P1}F. Canfora, A. Giacomini, J. Oliva, \textit{Phys. Rev.} \textbf{D
82}, (2010) 045014.

\bibitem {P2}A. Anabal\'{o}n, F. Canfora, A. Giacomini, J. Oliva,
\textit{Phys. Rev}. \textbf{D 83}, (2011) 064023.

\bibitem {P3}F. Canfora, A. Giacomini, J. Oliva, \textit{Phys. Rev.} \textbf{D
84}, (2011) 105019.

\bibitem {P4}M. de Cesare, G. Esposito H. Ghorbani, \textit{Phys. Rev.}
\textbf{D 88}, (2013) 087701.

\bibitem {SS05}R. F. Sobreiro, S. P. Sorella, "\textit{Introduction to the
Gribov Ambiguities In Euclidean Yang-Mills Theories}" arXiv:hep-th/0504095.

\bibitem {hedgehog}F. Canfora, P. Salgado-Rebolledo, \textit{Phys. Rev.}
\textbf{D} \textbf{87}, (2013) 045023.

\bibitem {DZ89}G. F. Dell'Antonio, D. Zwanziger, \textit{Nucl. Phys.}
\textbf{B 326}, (1989) 333.

\bibitem {Zwanziger-Ren}D. Zwanziger, \textit{Nucl. Phys.} \textbf{B323}
(1989) 513; D. Zwanziger, \textit{Nucl. Phys.} \textbf{B399} (1993) 477.

\bibitem {baal}P. van Baal, Nucl.Phys. B369 (1992) 25.

\bibitem {VZ}N. Vandersickel, D. Zwanziger, \textit{The Gribov Problem and QCD
Dynamics}, Phys. Rept. \textbf{520, }175 (2012) [arXiv:1202.1491 [hep-th]].

\bibitem {local1}M. Maggiore, M. Schaden,\textit{ Phys. Rev}\textbf{. D 50}
(1994) 6616.

\bibitem {local2}J. A. Gracey, \textit{JHEP}\textbf{ 0605} (2006) 052.

\bibitem {local3}D. Dudal, R. F. Sobreiro, S. P. Sorella, H. Verschelde,
\textit{Phys. Rev}\textbf{. D 72} (2005) 014016.

\bibitem {local4}D. Dudal, S. P. Sorella, N. Vandersickel, H.
Verschelde,\textit{ Phys. Rev}\textbf{. D 77} (2008) 071501.

\bibitem {Sorella:2011}S.~P.~Sorella, D.~Dudal, M.~S.~Guimaraes and
N.~Vandersickel,
PoS FACESQCD , 022 (2010) [arXiv:1102.0574 [hep-th]].

\bibitem {Dudal:2008}D.~Dudal, J.~A.~Gracey, S.~P.~Sorella, N.~Vandersickel
and H.~Verschelde,
\textit{Phys.\ Rev.\ D} \textbf{78} (2008) 065047 [arXiv:0806.4348 [hep-th]].

\bibitem {sorellaPRL}D. Dudal, M.S. Guimaraes, S.P. Sorella, \textit{Phys.
Rev. Lett.}\textbf{ 106}, 062003 (2011).

\bibitem {fabrizio}F. Canfora, L. Rosa, \textit{Phys. Rev}\textbf{. D 88}
(2013) 045025.

\bibitem {le bellac}M. Le Bellac, "\textit{Thermal Field Theory}", Cambridge
University Press, Cambridge, 2000.

\bibitem {yagi}K. Yagi (Author), T. Hatsuda (Author), Y. Miake
"\textit{Quark-Gluon Plasma: From Big Bang to Little Bang}", Cambridge
University Press, Cambridge, 2005.

\bibitem {Glendenning}N.~K.~Glendenning, \textit{Compact stars: Nuclear
physics, particle physics, and general relativity}, New York, USA: Springer (1997).

\bibitem {kapusta2}J.~Kapusta, B.~Muller, J.~Rafeslki, \textit{Quark-Gluon
Plasma: Theoretical Foundations}, Amsterdam, Netherlands: Elseiver (2003).

\bibitem {Korthals Altes}K. Altes, C.~P.\ 2004, \textit{Symmetries and
Quasi-Particles in Hot QCD, }Proceedings of the International School of
Subnuclear Physics. Edited by A. Zichichi. Singapore: World Scientific
Publishing, 2004.

\bibitem {gupta}S. Gupta, \textit{Phys. Rev. }\textbf{D 64} (2001) 034507.

\bibitem {Liao:2007}J. Liao, E. Shuryak \textit{Phys. Rev.} \textbf{C 75}
(2007) 054907, arXiv:0611131 [hep-ph].

\bibitem {Pisarski:2009}R. D. Pisarski, \textit{Nucl. Phys.} \textbf{B} (Proc.
Suppl.) 195 (2009) 157-198.

\bibitem {Hidaka:2008}Y.~Hidaka and R.~D.~Pisarski,
\textit{Phys.\ Rev.}\textbf{\ D 78}, 071501 (2008) arXiv:0803.0453 [hep-ph].

\bibitem {Kashiwa:2013}R.~D.~Pisarski, K.~Kashiwa and V.~Skokov,
\textit{Nucl.\ Phys}.\textbf{\ A} \textbf{904-905}, 973c (2013)
arXiv:1305.0767 [hep-ph].

\bibitem {zwanziger}D. Zwanziger, \textit{Phys.Rev}. \textbf{D 76}, (2007) 12504.

\bibitem {fukushima}K. Fukushima, N. Su, \textit{Phys.Rev. }\textbf{D 88},
(2013) 076008.

\bibitem {arnold}P. B. Arnold, \textit{Int.J.Mod.Phys.} \textbf{E16} (2007) 255.

\bibitem {reinosa}U. Reinosa, J. Serreau, M. Tissier, N. Wschebor,
arXiv:1311.6116 [hep-th].

\bibitem {Alkofer:2000wg}R.~Alkofer and L.~von Smekal,
Phys.\ Rept.\ \textbf{353}, 281 (2001) [hep-ph/0007355].

\bibitem {Peskin}M.E. Peskin, D.V. Schroeder, \textit{An Introduction to
Quantum Field Theory}. Perseus Books Publishing L.L.C. (1995).

\bibitem {kapusta}J.I. Kapusta, C. Gale, "\textit{Finite-Temperature Field
Theory: Principles and Applications}", Cambridge University Press, Cambridge, 2006.

\bibitem {sorella dm}R.F. Sobreiro, S.P. Sorella, D. Dudal, H. Verschelde,
\textit{Phys.Lett.} \textbf{B 590 }(2004) 265-272.

\bibitem {huang}S. Huang, \textit{Nucl.Phys}. \textbf{B 438} (1995) 54.

\bibitem {prosperi}G.M. Prosperi, M. Raciti, C. Simolo,
\textit{Prog.Part.Nucl.Phys.} 58 (2007) 387.

\bibitem {higgs1}M. A. L. Capri, D. Dudal, A. J. Gomez, M. S. Guimaraes, I. F.
Justo, S. P. Sorella, D. Vercauteren, \textit{ Phys. Rev}\textbf{. D 88}
(2013) 085022.

\bibitem {higgs2}M. A. L. Capri, D. Dudal, A. J. Gomez, M. S. Guimaraes, I. F.
Justo, S. P. Sorella, \textit{Eur. Phys. J. C} (2013) 73:2346.

\bibitem {frashe}E. Fradkin, S. Shenker, \textit{ Phys. Rev}\textbf{. D 19}
(1979) 3682.

\bibitem {kondo}K.-I. Kondo, "\textit{Decoupling and scaling solutions in
Yang-Mills theory with the Gribov horizon}", arXiv:0909.4866 [hep-th].

\bibitem {boucaud1}Ph. Boucaud, J.P. Leroy, A. Le Yaouanc, J. Micheli, O. Pen
e and J. Rodriguez-Quintero, \textit{JHEP}\textbf{06, 099} (2008), arXiv:0803.2161[hep-ph].

\bibitem {boucaud2}Ph. Boucaud, J.P. Leroy, A. Le Yaouanc, J. Micheli, O. Pen
e and J. Rodriguez-Quintero, \textit{JHEP}\textbf{06, 012} (2008), arXiv:0801.2721[hep-ph].

\bibitem {cucchieri1}A. Cucchieri and T. Mendes, PoS \textbf{LAT2007} (2007) 297.

\bibitem {bogolusky}I. L. Bogolubsky, E. M. Ilgenfritz, M. Muller-Preusske r
and A. Sternbeck, PoS \textbf{LAT2007} (2007) 290.

\bibitem {cucchieri2}A. Cucchieri and T. Mendes, \textit{Phys. Rev. Lett.}
\textbf{100} (2008) 241601, arXiv:0712.3517 [hep-lat].

\bibitem {cucchieri3}A. Cucchieri and T. Mendes, \textit{Phys. Rev.} \textbf{D
78} (2008) 094503, arXiv:0804.2371 [hep-lat].

\bibitem {furui}S. Furui, H. Nakajima, \textit{Phys. Rev.} \textbf{D 73}
(2006) 074503.

\bibitem {parappilly}M. B. Parappilly, P. O. Bowman, U. M. Heller, D. B.
Leinweber, A. G. Williams, J. B. Zhang, \textit{Phys. Rev.} \textbf{D 73}
(2006) 054504.

\bibitem {burgio}G. Burgio, M. Schrock, H. Reinhardt, M. Quandt, \textit{Phys.
Rev.} \textbf{D 86} (2012) 014506.

\bibitem {rojas}E. Rojas, J. P. B. C. de Melo, B. El-Bennich, O. Oliveira, T.
Frederico, arXiv:1306.3022 [hep-ph].

\bibitem {sorella_quarks}D. Dudal, M.S. Guimaraes, L.F. Palhares, S.P.
Sorella, "\textit{From QCD to a dynamical quark model: construction and some
meson spectroscopy}", arXiv:1303.7134 [hep-ph].

\bibitem {benic}S. Beni{\'c}, D. Blaschke, M. Buballa, \textit{Phys. Rev.}
\textbf{D 86} (2012) 074002, arXiv:1206.6582 [hep-ph].
\end{thebibliography}
\end{document}